\documentstyle[11pt,twoside,epsfig,rotating]{article}

\newlength{\dinwidth}
\newlength{\dinmargin}
\begin{document}  
\newcommand{\rgfd}{\mbox{RG--${\rm F_2^D}$ (fit 3)}}
\newcommand{\rgqq}{\mbox{RG--${\rm F_2^D}$ (fit 1)}}
\newcommand{\lsim}{\raisebox{-0.5mm}{$\stackrel{<}{\scriptstyle{\sim}}$}}
\newcommand{\gsim}{\raisebox{-0.5mm}{$\stackrel{>}{\scriptstyle{\sim}}$}}
\newcommand{\pom}{{I\!\!P}}
\newcommand{\slowpi}{\pi_{\mathit{slow}}}
\newcommand{\gevsq}{\mathrm{GeV}^2}
\newcommand{\fiidiii}{F_2^{D(3)}}
\newcommand{\fiidiiiarg}{\fiidiii\,(\beta,\,Q^2,\,x)}
\newcommand{\n}{1.19\pm 0.06 (stat.) \pm0.07 (syst.)}
\newcommand{\nz}{1.30\pm 0.08 (stat.)^{+0.08}_{-0.14} (syst.)}
\newcommand{\fiidiiiful}{F_2^{D(4)}\,(\beta,\,Q^2,\,x,\,t)}
\newcommand{\fiipom}{\tilde F_2^D}
\newcommand{\ALPHA}{1.10\pm0.03 (stat.) \pm0.04 (syst.)}
\newcommand{\ALPHAZ}{1.15\pm0.04 (stat.)^{+0.04}_{-0.07} (syst.)}
\newcommand{\fiipomarg}{\fiipom\,(\beta,\,Q^2)}
\newcommand{\pomflux}{f_{\pom / p}}
\newcommand{\nxpom}{1.19\pm 0.06 (stat.) \pm0.07 (syst.)}
\newcommand {\gapprox}
   {\raisebox{-0.7ex}{$\stackrel {\textstyle>}{\sim}$}}
\newcommand {\lapprox}
   {\raisebox{-0.7ex}{$\stackrel {\textstyle<}{\sim}$}}
\newcommand{\pomfluxarg}{f_{\pom / p}\,(x_\pom)}
\newcommand{\dsf}{\mbox{$F_2^{D(3)}$}}
\newcommand{\dsfva}{\mbox{$F_2^{D(3)}(\beta,Q^2,x_{I\!\!P})$}}
\newcommand{\dsfvb}{\mbox{$F_2^{D(3)}(\beta,Q^2,x)$}}
\newcommand{\dsfpom}{$F_2^{I\!\!P}$}
\newcommand{\gap}{\stackrel{>}{\sim}}
\newcommand{\lap}{\stackrel{<}{\sim}}
\newcommand{\fem}{$F_2^{em}$}
\newcommand{\tsnmp}{$\tilde{\sigma}_{NC}(e^{\mp})$}
\newcommand{\tsnm}{$\tilde{\sigma}_{NC}(e^-)$}
\newcommand{\tsnp}{$\tilde{\sigma}_{NC}(e^+)$}
\newcommand{\st}{$\star$}
\newcommand{\sst}{$\star \star$}
\newcommand{\ssst}{$\star \star \star$}
\newcommand{\sssst}{$\star \star \star \star$}
\newcommand{\tw}{\theta_W}
\newcommand{\sw}{\sin{\theta_W}}
\newcommand{\cw}{\cos{\theta_W}}
\newcommand{\sww}{\sin^2{\theta_W}}
\newcommand{\cww}{\cos^2{\theta_W}}
\newcommand{\trm}{m_{\perp}}
\newcommand{\trp}{p_{\perp}}
\newcommand{\trmm}{m_{\perp}^2}
\newcommand{\trpp}{p_{\perp}^2}
\newcommand{\alp}{\alpha}
\newcommand{\alps}{\alpha_s}
\newcommand{\sqrts}{$\sqrt{s}$}
\newcommand{\LO}{$O(\alpha_s^0)$}
\newcommand{\Oa}{$O(\alpha_s)$}
\newcommand{\Oaa}{$O(\alpha_s^2)$}
\newcommand{\PT}{p_{\perp}}
\newcommand{\JPSI}{J/\psi}
\newcommand{\GP}{\gamma p  \rightarrow }
\newcommand{\Gg}{\gamma g  \rightarrow }
\newcommand{\sh}{\hat{s}}
\newcommand{\uh}{\hat{u}}
\newcommand{\MP}{m_{J/\psi}}
\newcommand{\PO}{I\!\!P}
\newcommand{\xbj}{x}
\newcommand{\xpom}{x_\pom}
\newcommand{\ttbs}{\char'134}
\newcommand{\AmS}{{\protect\the\textfont2
  A\kern-.1667em\lower.5ex\hbox{M}\kern-.125emS}}
\newcommand{\xpomlo}{3\times10^{-4}}  
\newcommand{\xpomup}{0.05}
\hyphenation{sea-gull}
\begin{titlepage}
\begin{flushleft}
{\tt DESY 98-029}\hfill {\tt ISSN 0418-9833} \\
\end{flushleft}
\vspace*{3.0cm}
\begin{center}\begin{LARGE}
{\bf Hadron Production in Diffractive Deep-Inelastic Scattering}\\
\vspace*{2.5cm}
H1 Collaboration \\
\vspace*{2.5cm}
\end{LARGE}
{\bf Abstract}
\begin{quotation}
\noindent
Characteristics of hadron production in diffractive deep-inelastic
positron-proton scattering are studied using data collected in $1994$
by the H1 experiment at HERA\@. 
The following distributions are measured in the centre-of-mass frame
of the photon dissociation system:
the hadronic energy flow,
the Feynman-$x$ ($x_F$) variable for charged particles,
the squared transverse momentum of charged particles ($p_T^{*2}$),
and the mean $p_T^{*2}$ as a function of $x_F$.
These distributions are compared with results in the $\gamma^* p$
centre-of-mass frame from inclusive 
deep-inelastic scattering in the fixed-target experiment EMC, 
and also with the predictions of several Monte Carlo calculations.
The data are consistent with a picture in which the partonic structure
of the diffractive exchange is dominated at low $Q^2$ by hard gluons.

\end{quotation}
\vspace*{2.0cm}
{\it Submitted to Physics Letters $\mbox{\boldmath $B$}$}  \\
\vfill
\cleardoublepage
\end{center}
\end{titlepage}
\noindent
\begin{flushleft}
 C.~Adloff$^{35}$,                
 S.~Aid$^{13}$,                   
 M.~Anderson$^{23}$,              
 V.~Andreev$^{26}$,               
 B.~Andrieu$^{29}$,               
 V.~Arkadov$^{36}$,               
 C.~Arndt$^{11}$,                 
 I.~Ayyaz$^{30}$,                 
 A.~Babaev$^{25}$,                
 J.~B\"ahr$^{36}$
% ...

\end{flushleft}
\begin{flushleft}
\noindent
% ...
$ ^5$ Rutherford Appleton Laboratory, Chilton, Didcot, UK$^b$ \\
% ...
$ ^7$ Physics Department and IIRPA,
       University of California, Davis, California, USA$^e$ \\
% ...
$ ^{11}$ DESY, Hamburg, Germany$^a$ \\
% ...
$ ^{13}$ II. Institut f\"ur Experimentalphysik, Universit\"at Hamburg,
          Hamburg, Germany$^a$  \\
% ...
$ ^{23}$ Department of Physics and Astronomy, 
          University of Manchester, Manchester, UK$^b$ \\
% ...
$ ^{25}$ Institute for Theoretical and Experimental Physics,
          Moscow, Russia \\
 $ ^{26}$ Lebedev Physical Institute, Moscow, Russia$^{f,k}$ \\
% ...
$ ^{29}$ LPNHE, Ecole Polytechnique, IN2P3-CNRS, Palaiseau, France \\
 $ ^{30}$ LPNHE, Universit\'{e}s Paris VI and VII, IN2P3-CNRS,
          Paris, France \\
% ...
$ ^{35}$ Fachbereich Physik, Bergische Universit\"at Gesamthochschule
          Wuppertal, Wuppertal, Germany$^a$ \\
 $ ^{36}$ DESY, Institut f\"ur Hochenergiephysik, Zeuthen, Germany$^a$ \\
% ...
$ ^a$ Supported by the Bundesministerium f\"ur Bildung, Wissenschaft,
        Forschung und Technologie, FRG,
        under contract numbers 7AC17P, 7AC47P, 7DO55P, 7HH17I, 7HH27P,
        7HD17P, 7HD27P, 7KI17I, 6MP17I and 7WT87P \\
 $ ^b$ Supported by the UK Particle Physics and Astronomy Research
       Council, and formerly by the UK Science and Engineering Research
       Council \\
% ...
$ ^e$ Supported in part by US~DOE grant DE~F603~91ER40674 \\
 $ ^f$ Supported by the Deutsche Forschungsgemeinschaft \\
% ...
$ ^k$ Supported by Russian Foundation for Basic Researches 
       grant no. 96-02-00019 \\

\end{flushleft}
%
\newpage

\section{Introduction}
\label{sec:intro}

\begin{figure}[htb]
  \begin{center}
    \epsfig{file=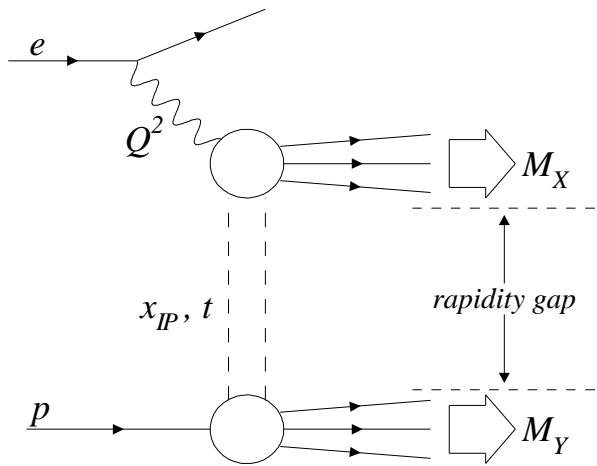,
            width=0.5\textwidth,
            bburx=405,bbury=345,bbllx=50,bblly=65,clip=}
    \caption{\footnotesize Schematic picture of 
              a deep-inelastic interaction  
              $ep\rightarrow eXY$ with a ``rapidity gap" 
              devoid of hadronic energy.
              The high-$Q^2$ virtual photon interacts 
              with a colourless space-like entity of squared  
              4-momentum $t$, whose longitudinal momentum as a 
              fraction of that of the target proton is $x_{\pom}$;
              the direction of the system $Y$
              is close to that of the target proton.}
      \label{fig:diffpicky}
   \end{center}
\end{figure}

Studies of the 1992 deep-inelastic electron-proton scattering events (DIS)
at HERA~\cite{rapgap-evts} revealed the presence of ``rapidity-gap 
events'' -- events of the form $ep \rightarrow eXY$ in which the
hadronic final state consists of two parts, $X$ and $Y$, separated
by a large region in pseudorapidity in which no hadrons are observed.
This is illustrated in figure~\ref{fig:diffpicky}.
The masses $M_X$ and $M_Y$ of these two systems, separated by the largest
rapidity gap in the event, are thus small compared to
$W$, the invariant mass of the $\gamma^* p$ system.
The system $Y$ in these events consists of a proton or other low-mass 
hadronic state and has a momentum similar to that of the incoming proton. 
The magnitude of the square of the 4-momentum transferred to $X$
from the proton, $|t|$, is small ($\lsim\ 1\,\gevsq$).
The contribution of these events to the $ep$ interaction cross section 
has been measured in terms of a diffractive structure function, $\fiidiii$,
using the HERA data of 1993~\cite{f2d93}.
The more precise $\fiidiii$ measurements made with the 1994 data~\cite{H1QCD1}
indicate that the cross section for rapidity-gap events may
be parameterised as a diffractive contribution, from pomeron ($\pom$)
exchange, together with a contribution from meson exchange.
The pomeron contribution dominates for small values ($\lsim\ 0.05$) 
of the variable $\xpom$, which is the fraction of the incoming 
proton's longitudinal momentum carried by the exchange.
Furthermore, the pomeron may be interpreted as having partonic structure.

A QCD study of parton distribution functions~\cite{H1QCD1},
evolved according to the DGLAP equations~\cite{DGLAP},
reveals the preference of the $\fiidiii$ data for a pomeron that is
dominated by a ``hard-gluon'' parton distribution at the starting scale of
$Q^2_0=3\,$GeV$^2$ (fits~2 and~3 in~\cite{H1QCD1}), 
i.e.\ a distribution with a large contribution from gluons
carrying a significant fraction of the momentum of the pomeron.
A pomeron model with only quarks at $Q^2_0$ (fit 1 in~\cite{H1QCD1}) fails. 

The studies of the diffractive final state presented here may be
regarded as specific tests of the pomeron structure extracted from
$\fiidiii$, complementary to other final-state 
analyses~\cite{zeus:shape,h1:thrust,h1:jets}.
A gluon-dominated pomeron is expected to interact largely by
boson-gluon fusion (BGF); the transverse momenta of the outgoing partons 
relative to the photon direction in the
$\gamma^* \pom$ centre-of-mass (CM) frame (i.e.\ the CM frame of the
photon dissociation system $X$), as well as the amount of gluon radiation 
and consequently the parton multiplicity, are therefore likely to be greater
than in the case of a quark-dominated pomeron.  

The partonic structure of the pomeron should thus be reflected in the
energy-flow and charged-particle distributions in the 
$\gamma^* \pom$ CM frame.  
These distributions are compared here with data on deep-inelastic
$\mu p$ interactions in the $\gamma^* p$ CM frame at the EMC experiment.
This provides a comparison of the structure of the pomeron
with that of the proton.

The distributions are also compared with predictions
from a calculation using a pomeron with parton distributions from the 
best QCD fit to $\fiidiii$ (fit 3).  
Comparison with predictions from a calculation with only
quarks in the pomeron at the starting scale of the QCD fit (fit 1)
demonstrates the sensitivity
of the measurements to the parton distributions of the pomeron.

Another approach to diffractive DIS is provided by
the photon dissociation picture~\cite{phot-diss}.
In the rest frame of the proton, the photon fluctuates a long time 
before the interaction into a Fock state with definite parton content 
($q\bar{q}$, $q\bar{q}g$, \ldots),
and this partonic state scatters diffractively off the proton.
Scattering from a quark in the pomeron corresponds to an 
Aligned Jet Model~\cite{ajm} topology in this picture, whereas 
scattering from a gluon corresponds to Fock states with one or more 
gluons and results in less pronounced alignment.

A non-diffractive model for the production of events with a large
rapidity gap has also been proposed, using a conventional picture of
DIS based on scattering from a single parton within the proton,
followed by soft colour interactions~\cite{sci,mc:lepto}.
The data are also compared with predictions of this model.

%
%
\section{Detector, Event Selection and Kinematics \label{sec:sel}}
%
The data used here were collected in 1994 with the H1 detector
at the HERA collider, which operated with 27.5\,GeV positron and 
820\,GeV proton beams.
The H1 detector is described in detail elsewhere~\cite{H1-det};
those components of importance for the analyses presented here  are briefly 
mentioned in the following. 
The laboratory coordinate system has its origin at the nominal
interaction point and its $z$ axis in the proton beam direction,
also called the forward direction. 
The pseudorapidity is defined as $\eta = -\ln\tan\frac{\theta}{2}$, where 
$\theta$ is the angle with respect to the $z$ axis.

In the forward and central regions the interaction point is surrounded by 
a system of tracking detectors -- interleaved drift and multi-wire
proportional chambers -- which cover the pseudorapidity range
$-1.5 < \eta < 2.8$ and the full azimuth.
The momenta of charged particles are determined for this analysis 
from their track 
curvature in the central jet chamber (CJC) in the uniform magnetic field 
of strength 1.15\,T generated by a superconducting solenoid. 
The CJC has a resolution of 
$\sigma_{p_T}/p_T \approx 0.006 p_T \oplus 0.008$ (where $p_T$ is in GeV
and the constant term describes the contribution from multiple
scattering at $p_T = 0.5$\,GeV)
and $\sigma_{\theta} \approx 20$\,mrad.
The solenoid surrounds the trackers and the fine-grained liquid-argon (LAr) 
calorimeter, which covers the range $ -1.5 < \eta < 3.6$ and is used to 
measure energies in the hadronic final state with a resolution of
$\sigma_E/E\approx 0.5/\sqrt{E} \oplus 0.02$ 
(where $E$ is in GeV)~\cite{H1-det}. 
Charged-particle detection in the backward region, $-3.1 < \eta < -1.5$, 
is provided by the backward proportional chamber (BPC)\@. 
Behind this, the backward electromagnetic calorimeter (BEMC)
completes the calorimetric coverage in the range $-3.5 < \eta < -1.5$. 

Several subdetectors in the forward region are used
in this analysis to tag particles emitted close to the proton direction.
These detectors cover a range larger than their purely geometrical acceptance,
due to the effect of secondaries resulting from the scattering of
primary particles in the beam pipe and adjacent material.
The copper-silicon plug calorimeter covers the range $3.5 < \eta < 5.5$.  
The first three double layers of drift chambers
in the forward muon detector (FMD) are sensitive to particles in the
range $5.0 < \eta < 6.5$.  The proton-remnant tagger (PRT), comprising a
set of double layers of scintillators situated around the proton beam pipe,
covers the range $6.0 < \eta < 7.5$.

The luminosity is determined from the rate of Bethe-Heitler interactions
detected in a monitor downstream of the main detector in the
positron direction.
The total luminosity analysed here amounts to approximately $2\,$pb$^{-1}$.

All events included in this analysis are required to have a scattered
positron reconstructed with an energy greater than 12\,GeV, emitted
in the angular range $156^{\circ} < \theta < 173^{\circ}$.
The scattered positron is identified by looking for the highest-energy
cluster in the BEMC\@.  This cluster is required to be associated with
a hit in the BPC and to pass a cut on its transverse size.
The trigger used to select these events for read-out requires the
presence of an energy deposit of at least 4\,GeV in the BEMC\@.
The efficiency of the BEMC trigger for positron energies larger than
10\,GeV is known to exceed $99\%$~\cite{triggeff}. For events passing the
above selection the photoproduction background is
less than $1\%$.

The data are selected to lie in the kinematic region 
$7.5 < Q^2 < 100\,$GeV$^2$ and $0.05 < y < 0.6$.
These variables are defined as $Q^2 = -q^2$ and 
$y = (q \cdot p) / (k \cdot p) \simeq (W^2+Q^2)/s$, where
$p$, $k$ and $q$ are the 4-momenta of the incoming proton, positron
and photon respectively, $W$ is the CM energy of the $\gamma^* p$
system, and $s$ is the squared CM energy of the $ep$ system.
The variables $Q^2$ and $y$ are calculated from the measured energy 
and angle of the scattered positron.

Diffractive events are recognised by the rapidity gap between the
outgoing hadronic systems $X$ and $Y$\@.
This gap is tagged by the absence of signals above noise in the forward 
components of the H1 detector~\cite{H1QCD1,thesis:mehta}: 
the plug calorimeter, the FMD and the PRT\@.
There must also be no energy deposit of more than 400\,MeV in the
forward region ($\eta>3.0$) of the LAr calorimeter.

The variable $\xpom$ is defined as
$$
x_{\pom} = \frac{q \cdot \pom}{q \cdot p} 
$$
where $\pom$ is the 4-momentum exchanged between systems $X$ and $Y$,
and is reconstructed using the relation
$$
x_{\pom} \simeq \frac{1}{2E_p} \sum_{e+X}{(E+p_z)}.
$$
Here, $E_p$ is the energy of the incoming proton, $E$ and $p_z$ are
the energy and longitudinal momentum of each final-state particle
in the laboratory frame, and the sum runs over the scattered positron $e$
and all detected particles in the photon dissociation system $X$\@.
The particles are reconstructed using a combination of tracks
and calorimeter clusters, with an algorithm for 
track-cluster association avoiding double counting~\cite{h1:diff-photo}.
In order to enhance the contribution from pomeron exchange and to reduce
the contribution from meson exchange, the requirement
$\xpom < 0.025$ is made.
The fraction of the exchanged momentum $\pom$ carried by the 
struck quark is given by $\beta$, where 
$$\beta = \frac{Q^2}{2q \cdot \pom} \simeq \frac{Q^2}{Q^2+M_X^2}.$$

Neither the squared momentum transfer $t$ nor the mass $M_Y$ of the 
hadronic system $Y$ is measured here.  However, the requirement 
of the absence of activity in the forward detectors imposes the 
approximate restrictions $M_Y < 1.6\,$GeV and $|t| < 1\,$GeV$^2$, 
and the results are corrected to this kinematic region.      
Since $M_Y$ is not measured, it is not possible to distinguish events 
containing an elastically scattered proton from those in which the proton 
dissociates into a low-mass state.

The analyses are performed in the $\gamma^{*} \pom$ CM frame,
the 4-momentum of the $\gamma^{*} \pom$ system being reconstructed as
$V = q + \xpom p$~\cite{thesis:hfs}.
The transverse momentum of the pomeron with respect to the direction
of the incoming proton is not measured, but can be neglected 
since this has no significant effect on the measurements presented here.

The energy-flow distribution is measured using energy deposits
reconstructed from clusters of cells in the LAr and BEMC\@. The clusters
are required to have an energy in the range $0 < E < 25 \, {\rm GeV}$ and
to lie in the angular range $-3 < \eta < 4$.
The charged-particle distributions are measured using only particles 
that are detected in the CJC, originate from the $ep$ interaction vertex
in the angular region $-1.31 < \eta < 1.31$, and have a transverse momentum
in the laboratory frame of $0.15 < p_T< 10 \, {\rm GeV}$, 
but the results are 
fully corrected with Monte Carlo simulations (see below)
to cover the whole hadronic system. 

Two variables are calculated for each charged particle, both
defined in the CM frame of the $\gamma^{*} \pom$ system.
The Feynman-$x$ variable, $x_F$, is defined as
$$x_F=\frac{2p^*_{\parallel}}{M_X},$$
where $p^*_{\parallel}$ is the component of the particle's momentum in the
direction of motion of the incoming photon.
In non-diffractive DIS, $x_F$ is defined in the $\gamma^{*} p$ CM frame, 
and the denominator is $W$\@.
Positive $x_F$ corresponds to the current hemisphere and negative $x_F$
to the target fragmentation hemisphere.
The other variable used here, $p_T^{*}$, is the transverse momentum of the 
particle with respect to the photon direction.

%
%
%
\section{Monte Carlo Models for Diffractive Interactions \label{sec_mc}}
%
Monte Carlo calculations employing the event generator
RAPGAP~2.02~\cite{rapgap} 
are used in conjunction with a detailed simulation of the H1 apparatus to 
correct the experimental distributions for the acceptance and resolution 
of the detector, and later to compare the results with theoretical 
expectations.  The data are also compared with the predictions of the 
Monte Carlo generator LEPTO~6.5~\cite{mc:lepto}.

The RAPGAP model treats diffractive interactions as inelastic $e\pom$ 
collisions, the pomeron being modelled as an object with partonic 
substructure.
Scattering on mesons is also included.

\begin{figure}[htb] \unitlength 1cm
      \begin{center}
         \begin{picture}(14.5,6.5)
          \put(0.,0.){\epsfig{
             figure=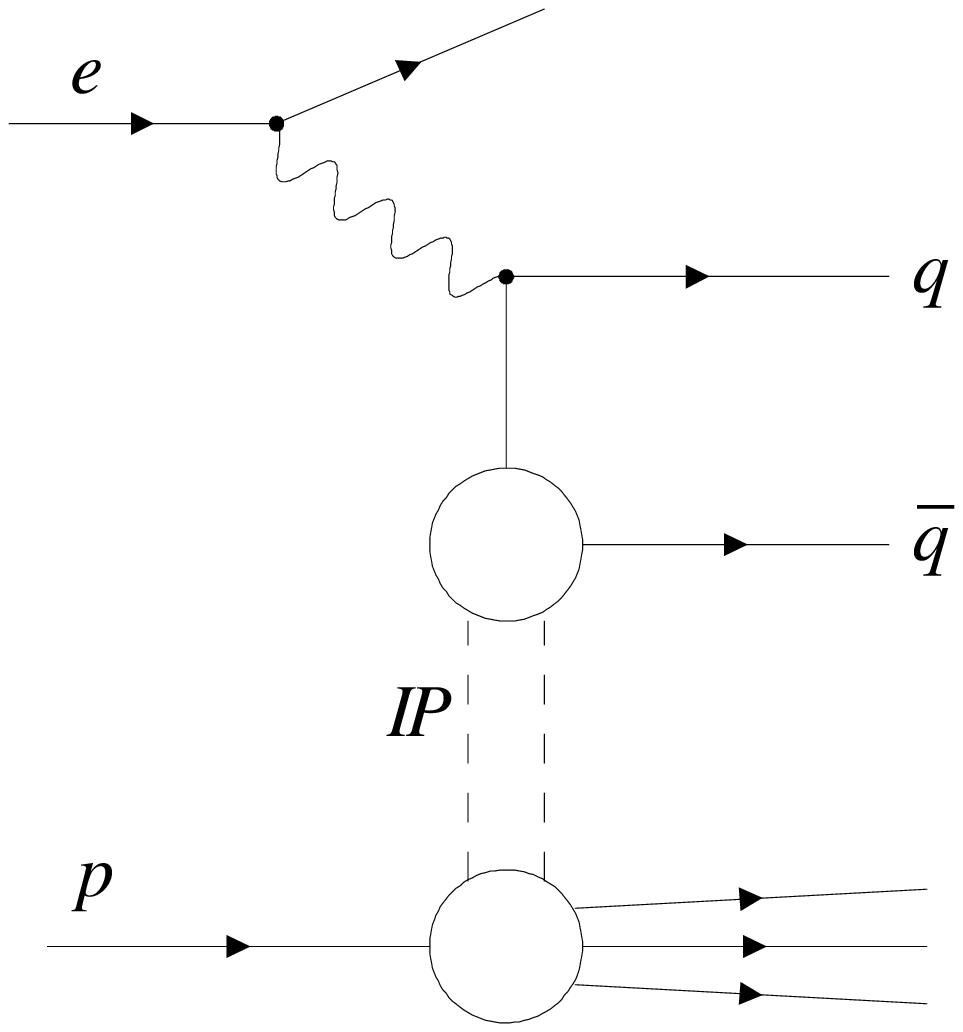,
             width=4.5cm,bburx=380,bbury=412,bbllx=95,bblly=112,clip=}}
          \put(5.,0.){\epsfig{
             figure=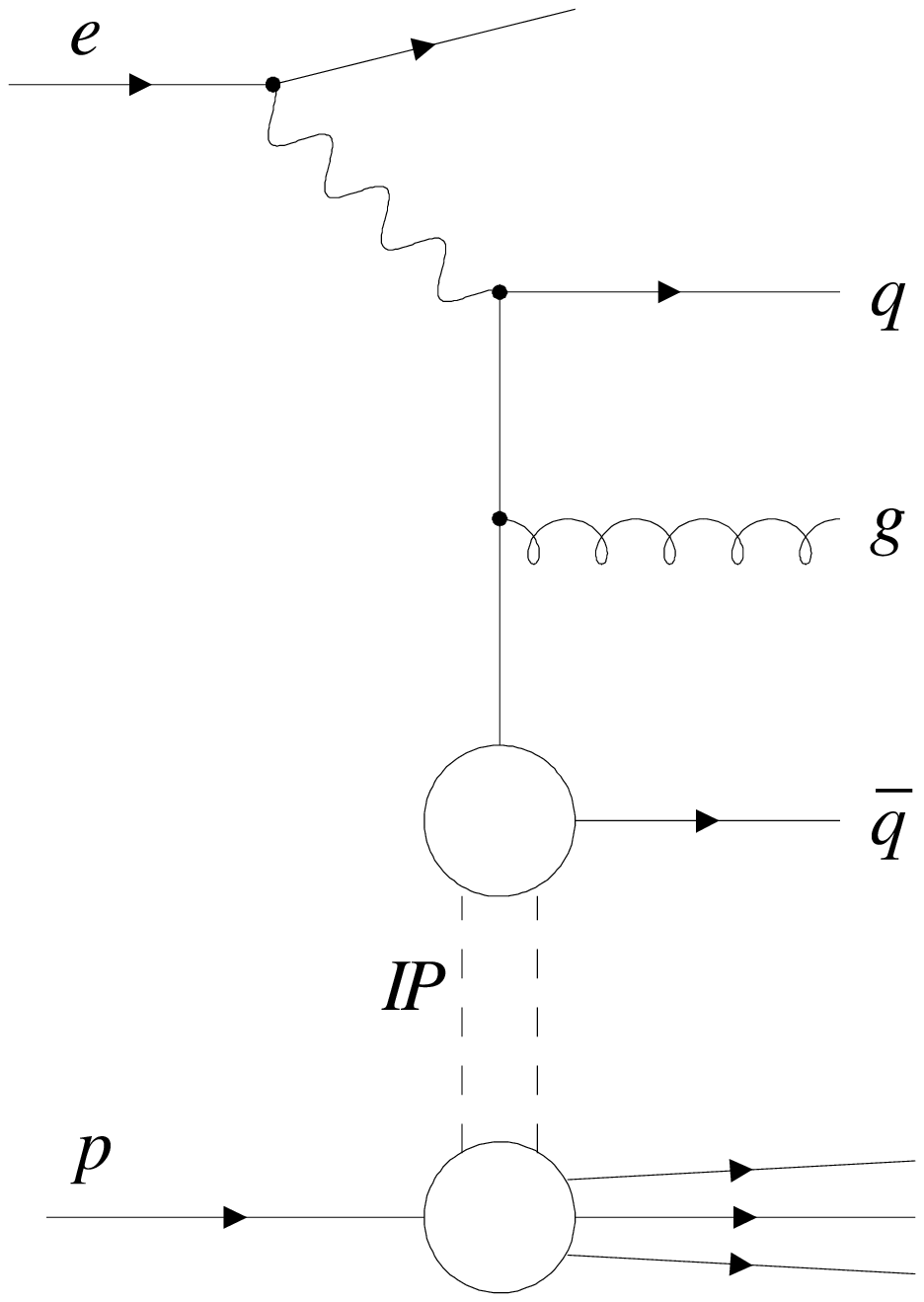,
             width=4.5cm,bburx=370,bbury=450,bbllx=95,bblly=70,clip=}}
          \put(10.,0.){\epsfig{
             figure=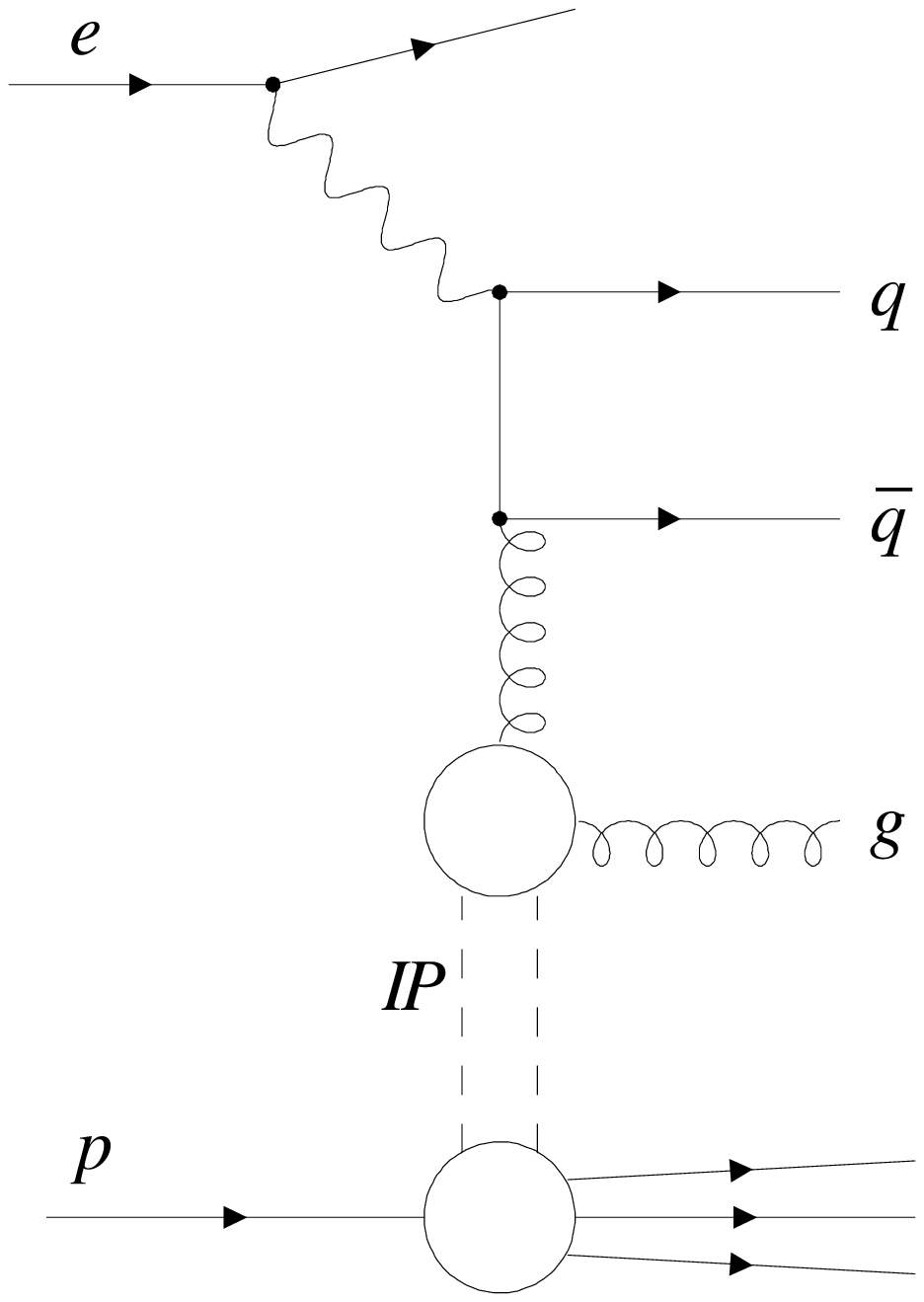,
             width=4.5cm,bburx=370,bbury=450,bbllx=95,bblly=70,clip=}}
          \put(0,5.){(a)}
          \put(5,5.){(b)}
          \put(10,5.){(c)}
         \end{picture}
         \caption{\footnotesize Elementary processes included in the
                  RAPGAP simulation of $ep$ diffractive interactions:
                  (a) lowest-order, ${\cal O}(\alpha)$, $e$ scattering from a 
                  quark, leading to a two-parton configuration;
                  (b) one of the diagrams for the
                  QCD-C process, ${\cal O}(\alpha\alpha_s)$;
                  (c) lowest-order, ${\cal O}(\alpha\alpha_s)$, $e$ 
                  scattering from a gluon by BGF, leading to a three-parton
                  configuration.}
             \label{fig:motiv}
      \end{center}
  \end{figure}

For scattering from quarks, the lowest-order diagram considered by RAPGAP
at the parton level is ${\cal O}(\alpha)$~$eq \rightarrow eq$ scattering
(figure~\ref{fig:motiv}a); a higher-order, ${\cal O}(\alpha\alpha_s)$,
process is QCD Compton (QCD-C) scattering \mbox{$eq  \rightarrow eqg$}
(figure~\ref{fig:motiv}b).
For scattering from gluons, 
the lowest-order ${\cal O}(\alpha \alpha_s)$ process is boson-gluon fusion 
$eg \rightarrow e q\bar{q}$ (figure~\ref{fig:motiv}c).
A cut ${\hat p}_T^{2}>2\,$GeV$^2$ is applied in order to avoid 
divergences in the ${\cal O}(\alpha \alpha_s)$ matrix elements for 
massless quarks, 
${\hat p}_T$ being the transverse momentum of the outgoing partons
with respect to the photon direction in the CM frame of the hard subprocess.
For ${\hat p}_T^{2}<2\,$GeV$^2$, only $eq \rightarrow eq$ scattering is used.
In the RAPGAP model, a ``pomeron remnant'' is implicit at the parton level,
consisting of a quark (antiquark) in $eq$ ($e \bar{q}$)
scattering and a gluon in the $eg$ case (see figure~\ref{fig:motiv}).
The renormalisation and factorisation scale $\mu^2$ is set to $Q^2$.
Higher-order corrections at the parton level are treated using
leading-log parton showers (PS)~\cite{MEPS} as implemented in RAPGAP\@.
Subsequent hadronisation is simulated according to the Lund string
model in JETSET~\cite{JETSET}\@.
The dependence of the acceptance corrections on the method used to 
handle higher-order corrections in RAPGAP is studied with a 
separate calculation in which QCD-C and higher-order processes are simulated 
by the colour-dipole (CD) model~\cite{ARIADNE}.
RAPGAP is interfaced to HERACLES~\cite{HERACLES} for the simulation
of QED radiative effects.

For the comparisons presented here, two sets of parton distributions for the
pomeron are used in the RAPGAP generator.
The first is taken from the best fit to the $Q^2$ and $\beta$
dependence of the diffractive structure function $\fiidiii$
(fit 3 in~\cite{H1QCD1}).
It is characterised by a hard gluon distribution at the starting scale 
for DGLAP evolution, \mbox{$Q_0^{2}=3\,{\rm GeV^{2}}$}, in which gluons carry
$\gsim\ 80\%$ of the total momentum of the diffractive 
exchange\footnote{Fits 2 and 3 in~\cite{H1QCD1} give very similar
predictions for the distributions studied here.  Only the predictions
from fit 3 are shown in this paper.}. 
The second is taken from a fit in which only quarks are permitted 
to contribute to the partonic structure of the pomeron at $Q^2=Q_0^2$
(fit 1 in~\cite{H1QCD1}). 
The latter choice does not provide a satisfactory description of the 
$F_2^{D(3)}$ measurements; it is used here to demonstrate the sensitivity
of the results to the parton distributions of the pomeron.
The predictions of RAPGAP with these two sets of parton distributions are
hereafter referred to as \rgfd\ and \rgqq\ respectively. 
The meson structure function is taken to be that of the pion~\cite{owens}.
In both cases, the parton distribution functions evolve with $Q^2$
according to the DGLAP equations.
The partons are treated as having no intrinsic $k_T$.

In the LEPTO~6.5 model, rapidity-gap events are generated by the 
soft colour interaction (SCI) mechanism.  Deep-inelastic
scattering is modelled using the matrix elements for processes up to
${\cal O}(\alpha\alpha_s)$, as in RAPGAP, but with partons coming
directly from the proton according to the MRS(H) 
parameterisation~\cite{mrsh} of the proton structure function,
in which the renormalisation and factorisation scale $\mu^2$ is
again set to $Q^2$.  
The divergences in the ${\cal O}(\alpha \alpha_s)$ matrix elements are
avoided using the cuts $\hat{s}>\hat{s}_{\mathit min}$ and
$z_q<z_{q,{\mathit min}}<1-z_q$, where $\hat{s}$ is the CM energy
of the hard subprocess, $z_q = (p \cdot p_q)/(p \cdot q)$ and
$p_q$ is the 4-momentum of one of the outgoing partons.
The parameters $\hat{s}_{\mathit min}$ and $z_{q,{\mathit min}}$
are set to $4\,\gevsq$ and 0.04 respectively.
Higher-order corrections are treated
with the PS method, and hadronisation follows the Lund string model.
Further non-perturbative interactions take place as the outgoing
partons pass through the colour field of the proton.
These soft colour interactions can result in a 
hadronic final state comprising two colour-singlet systems separated 
by a rapidity gap.

%
%
%
\section{Acceptance Corrections and Systematic Errors
         \label{sec_syst}}

The data are corrected for the acceptance and resolution of the H1
apparatus using events generated by RAPGAP~2.02~\cite{rapgap} with
a hard-gluon pomeron structure function.
The following sources of systematic error are taken into account; the 
errors shown in brackets are those on the energy flow and the 
charged particle distributions respectively:
\vspace{-0.3cm}
\begin{itemize}
\item [$-$] an uncertainty in the LAr calibration for hadronic energy 
          of $\pm 4\%$, leading to an average error in the
          experimental distributions of $\pm (6\%,2\%)$;
\item [$-$] an uncertainty in the BEMC calibration for hadronic energy of 
          $\pm 20\%$, leading to an average error of $\pm (6\%, 0.7\%)$;
\begin{sloppypar}
\item [$-$] an uncertainty of $\pm 1\%$ in the positron energy scale,
          leading to an average error of $\pm (3\%, \, 0.5\%)$;
\end{sloppypar}
\item [$-$] an uncertainty of $\pm 1\,{\rm mrad}$ in the positron polar
          angle, leading to an average error of $\pm (0.5\%, \, 0.5\%)$;

\item [$-$] an uncertainty in the $\xpom$ dependence
          in the Monte Carlo used to calculate the acceptance corrections,
          taken into account by reweighting the generated events with
          the function $\xpom^{\pm 0.2}$, resulting in an average error
          of $\pm (8\%, 3\%)$;

\item [$-$] an uncertainty in the $\beta$ dependence in the Monte Carlo,
          taken into account by reweighting the generated events with
          the function $(a^{-1}-a)\beta+a$, where $a$ has the range
          0.5 to 2, resulting in an average error of $\pm (6\%,1\%)$;

\item [$-$] an uncertainty in the $t$ dependence in the Monte Carlo,
          taken into account by reweighting the generated events with
          the function $e^{\pm 2t }$,
          resulting in an average error of $\pm (5\%, 2\%)$;

\item [$-$] uncertainties arising from the method used to treat
          higher-order corrections in the Monte Carlo;
          this results in an error 
          of $\pm (10\%, 10\%)$, 
          evaluated by taking the full difference between
          the results obtained using the PS and CD models as an estimation 
          of the extra\-po\-lation uncertainty from the Monte Carlo.
\end{itemize}
In the charged-particle analysis, the track selection criteria were 
varied to allow for the imperfect description of the Central Tracker in the 
Monte Carlo. The selection was varied by increasing the angular range of the
tracks to $-1.50 < \eta < 1.64$ and by requiring the radial track length to
be greater than $15 \, {\rm cm}$ and the number of hits on a 
track $N_{hits}$ to be greater than 10.
This results in an average error of $\pm 5\%$,
rising in the low-$x_F$ region to $20\%$ in the $x_F$ distribution and
$15\%$ in $\langle p_T^{*2} \rangle$ in the ``seagull plot''
(figure~\ref{fig:seagull}).

%
%
%
\section{Results \label{sec_results}}

Results\footnote{The data are available in numerical form on request
and have also been submitted to the Durham HEPDATA database
{\tt http://durpdg.dur.ac.uk/HEPDATA}.}
are shown for the kinematic range $7.5<Q^2<100\,\gevsq$,
$0.05<y<0.6$, $\xpom<0.025$, $|t|<1\,\gevsq$ and $M_Y<1.6$\,GeV.
All distributions are fully corrected for the effects of the acceptance 
and resolution of the H1 apparatus, and statistical and systematic 
errors on the data points are combined quadratically.  
The inner vertical error bars indicate the statistical error and 
the outer bars the total error.
Due to the $x_\pom$ cut, the meson-exchange contribution is small
($< 7\%$ in the \rgfd\ model) and does not affect the conclusions. 

It is shown below that, in Monte Carlo calculations, the characteristic
features of the parton content of the pomeron are reflected 
in the distributions of various final-state observables.
The data are therefore compared with data on $\gamma^* p$ interactions
at relatively high Bjorken-$x$ values ($x > 0.01$),
where the structure of the proton is dominated by quarks.
The data used for this comparison are taken from $\mu p$ interactions
at the EMC experiment~\cite{EMC}, with the mean $\gamma^* p$ CM energy,
$\langle W \rangle = 14\,$GeV, of the EMC inclusive DIS data similar 
to the mean $\gamma^* \pom$ CM energy, $\langle M_X \rangle \approx 12$\,GeV, 
of the H1 diffractive data.

The experimental data are also compared with the predictions of the 
RAPGAP model with the two different pomeron structures from the
DGLAP fits to $\fiidiii$ -- the preferred hard-gluon structure (\rgfd) 
and the structure with only quarks at the starting scale (\rgqq) --
as well as with the soft colour interaction (SCI) model as 
implemented in LEPTO~6.5.

In the CM frame of the $\gamma^* \pom$ system, a model with a quark-dominated 
pomeron results in events in which the struck quark and the pomeron remnant
tend to be close to the $\gamma^*\pom$ axis.
Large contributions to the transverse momentum can arise from QCD 
corrections such as QCD Compton scattering, 
but the rate of these is suppressed by a factor $\alpha_s$.
In contrast, if the pomeron is a gluon-dominated object,
more transverse momentum and energy flow are produced at lowest order.
This arises because the quark propagator in the BGF process
can have non-zero virtuality, so the quark-antiquark pair from the hard 
subprocess is not necessarily aligned along the $\gamma^*\pom $ axis. 
Evidence for this effect is seen in a recent analysis, using the same data,
of the distribution of the $p_T$ of thrust
jets relative to the $\gamma^*\pom$ axis~\cite{h1:thrust}.
In both the quark- and gluon-dominated cases, 
small contributions to the transverse momentum are generated 
by the intrinsic $k_T$ of the partons, by the hadronisation process, and
by particle decays.
However, it is known from the average thrust distribution shown 
in~\cite{h1:thrust} that a quark-dominated exchange, even with
large intrinsic $k_T$,
would not be able to explain the relatively low value of
the average thrust.

\begin{figure}[tb]
\vspace*{-0.7cm}
  \begin{center}
    \epsfig{file=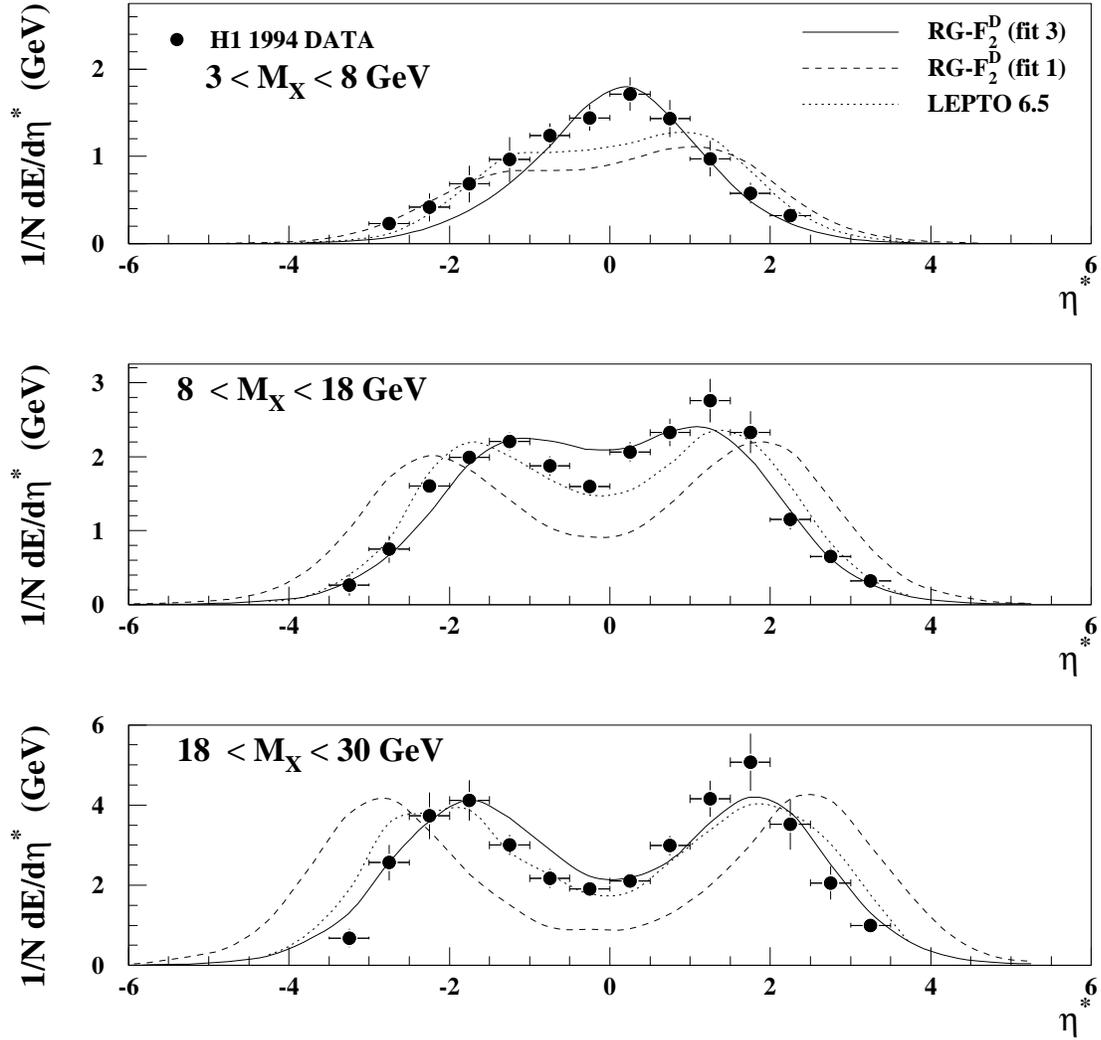,
            width=1.0\textwidth}
    \vspace*{-0.5cm}
    \caption{\footnotesize 
             The energy-flow distribution in 
             the $\gamma^* \pom$ CM frame for three different $M_X$
             intervals,
             in the kinematic region $7.5 < Q^{2} < 100\,$GeV$^2$, 
             $0.05 < y < 0.6$, $x_\pom < 0.025$, $|t|<1\,\gevsq$
             and $M_Y<1.6\,{\rm GeV}$.
             Positive $\eta^*$ corresponds to
             the direction of motion of the incoming photon.
             The statistical error bars are hidden by the symbols.
             The points are plotted at the centre of the bin in the
             horizontal coordinate, and the horizontal bars indicate the
             width of the bin.
             Also shown are the predictions of the following Monte Carlo 
             models: RAPGAP with the hard-gluon pomeron structure
             taken from fit~3 in~\protect\cite{H1QCD1} (\rgfd);
             RAPGAP with the pomeron structure containing
             only quarks at the starting scale, taken from fit 1 
             in~\protect\cite{H1QCD1} (\rgqq);
             and the soft colour interaction model as implemented in 
             LEPTO~6.5. The particles are taken as massless for this 
             distribution.}
    \label{fig:eflow}
\end{center}
\end{figure}

Figure~\ref{fig:eflow} shows the event-normalised energy flow 
$1 / N \  {\rm d}E / {\rm d}\eta^*$ in three different regions of $M_X$, 
$\eta^*$ being the pseudorapidity relative to the direction of 
motion of the incoming photon in the CM frame of the $\gamma^* \pom$ 
system and $N$ being the number of events.
The distribution is approximately symmetrical about $\eta^* = 0$, with
similar levels of energy flow in the two hemispheres.  
At higher masses, $M_X > 8$\,GeV, a two-peaked
structure is seen, indicating that the major topological property
of these events is of a 2-jet nature (current jet and remnant jet).

\begin{sloppypar}
The data are well described by the \rgfd\ model\footnote{
The ${\hat p}_T^{2}$ cut in the generator was varied from 2 up to
4\,$\gevsq$ without affecting the conclusions drawn from the
distributions presented in this paper.}, and for
$M_X > 8$\,GeV by LEPTO~6.5, whereas the \rgqq\ model predicts too much 
energy flow at the largest accessible pseudorapidity and fails
to account for the observed energy flow in the central region,
$\eta^{*}\approx0$.
\end{sloppypar}

These features can be related to the gluon content of the pomeron 
and to the role of BGF\@.
The transverse momentum generated by this process reduces the
accessible pseudorapidity range along the $\gamma^*\pom$ axis at fixed 
$M_X$ and induces additional energy flow in more central regions.
In addition, more energy is expected in the central region for a
gluon-dominated than for a quark-dominated object because of the exchange 
of a gluon in BGF; this results in a colour octet-octet field 
between the outgoing quarks and the gluonic remnant
(which all together form an overall colour singlet), giving
rise to enhanced soft gluon radiation compared to the triplet-antitriplet
field between the struck quark and the remnant in $eq$ 
scattering~\cite{thesis:hfs,dok:qcd}.
This effect has been seen for gluon jets
produced in $e^+e^-$ interactions~\cite{opal:gluons}. 
This enhancement is also apparent in the virtual photon dissociation
picture~\cite{phot-diss},
where the process corresponding to BGF involves a Fock state
containing a low-energy gluon.

The production of charged particles is studied using the distributions
of the variables $x_F$ and $p_T^{*2}$, defined in section~\ref{sec:sel}.
The results are shown for the restricted mass range
$8 < M_X < 18$\,GeV in order to make the data more directly comparable
with the EMC $\gamma^* p$ results.

\begin{figure}[tb] \unitlength 1cm
  \begin{center}
    \begin{picture}(15.,15.)
     \put(0.5,0.5){\epsfig
      {file= 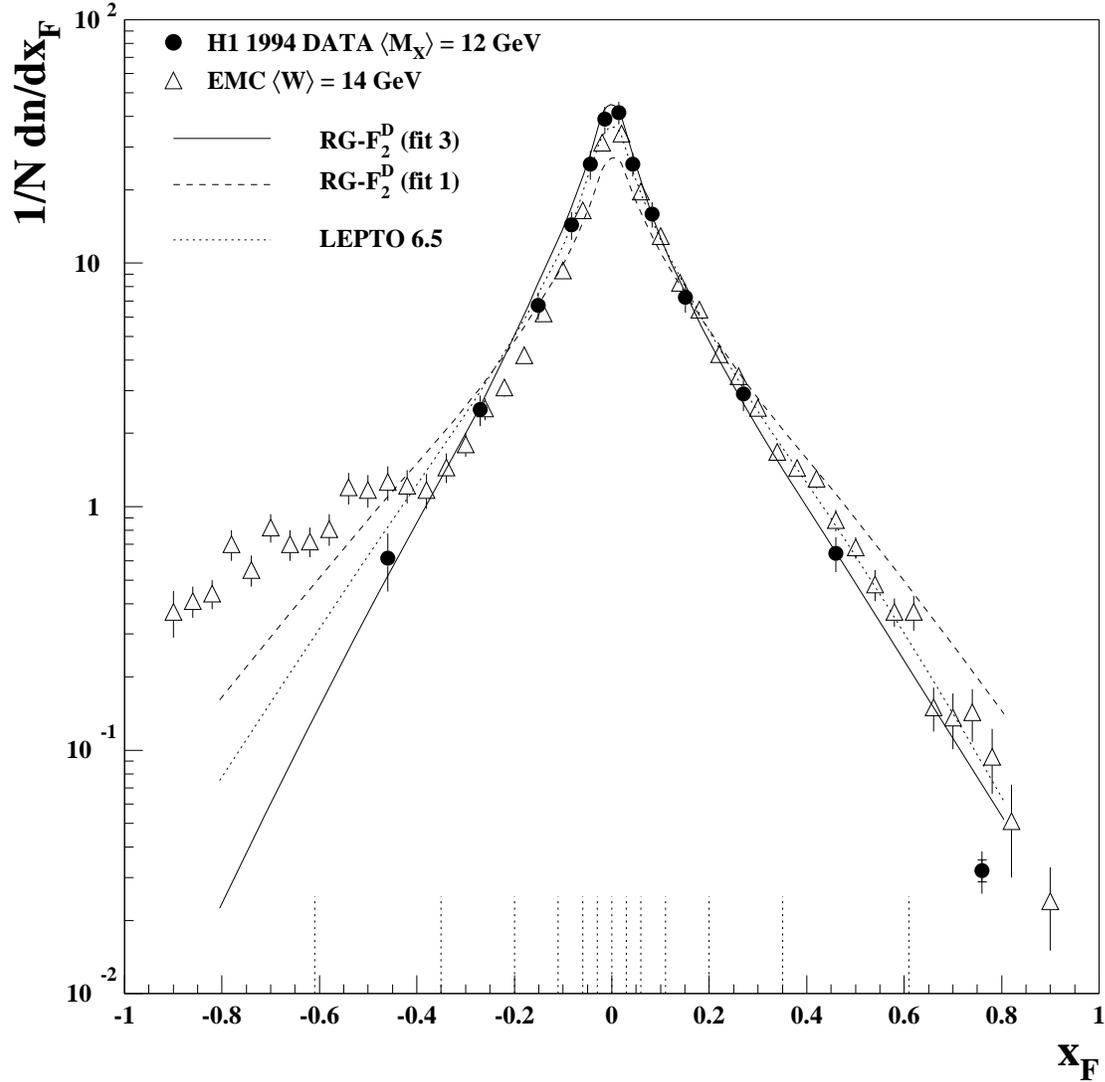,width=14cm,
           bburx=515,bbury=515,bbllx=25,bblly=40}}
    \end{picture}
    \caption{\footnotesize The Feynman-$x$ ($x_F$) distribution,
             showing 
             H1 $\gamma^*\pom$ data in the $\gamma^* \pom$ CM frame
             in the kinematic region
             $7.5 < Q^{2} < 100\,$GeV$^2$, 
             $0.05 < y < 0.6$, $\xpom < 0.025$, $8<M_X<18$\,GeV,
             $|t|<1\,\gevsq$ and $M_Y<1.6\,{\rm GeV}$,
             together with EMC
             $\mu p$ DIS data in the $\gamma^* p$ CM frame.
             Positive $x_F$ corresponds to the
             direction of motion of the incoming photon.
             Each point is plotted at the horizontal position where the
             predicted probability density function has its mean value 
             in the bin, as prescribed in~\protect\cite{stick}.
             The uncertainty in the horizontal position
             due to the model-dependence of the shape of the distribution 
             is indicated by horizontal bars, but in most cases is
             contained within the width of the symbol.
             The bin boundaries are marked by vertical dotted lines.
             Also shown are the predictions of the following Monte Carlo 
             models for the H1 data: 
             RAPGAP with the hard-gluon pomeron structure
             taken from fit~3 in~\protect\cite{H1QCD1} (\rgfd);
             RAPGAP with the pomeron structure containing
             only quarks at the starting scale, taken from fit 1 
             in~\protect\cite{H1QCD1} (\rgqq);
             and the soft colour interaction model as implemented in 
             LEPTO~6.5.}
    \label{fig:xf}
\end{center}
\end{figure}

The $x_F$ distribution, normalised by the number of events $N$, 
is shown in figure~\ref{fig:xf}.  
A feature of the diffractive data is the similarity 
between the $x_F$ distributions in the two hemispheres,
in contrast to the strong asymmetry of the $\gamma^{*}p$ data.
The asymmetry in the latter case is explained by 
the requirement of baryon number conservation
in $ep$ interactions.  This results in a large contribution from baryons
in the region \mbox{$x_F \, \lsim \, -0.4$}~\cite{EMC}.

The $x_F$ distribution is well described by the \rgfd\ and LEPTO~6.5
models, whereas the \rgqq\ calculation fails in that it predicts 
too little particle production in the central region ($x_F \approx 0$) 
and too much particle production at large $|x_F|$. 
These features are related to those observed in the 
energy-flow distribution: longitudinal phase space is restricted because of
the transverse momentum generated by the BGF process, and there is additional
radiation in the central region for a gluon-dominated object.

\begin{figure}[tb] \unitlength 1cm
  \begin{center}
    \begin{picture}(15.,15.)
     \put(0.5,0.5){\epsfig{file=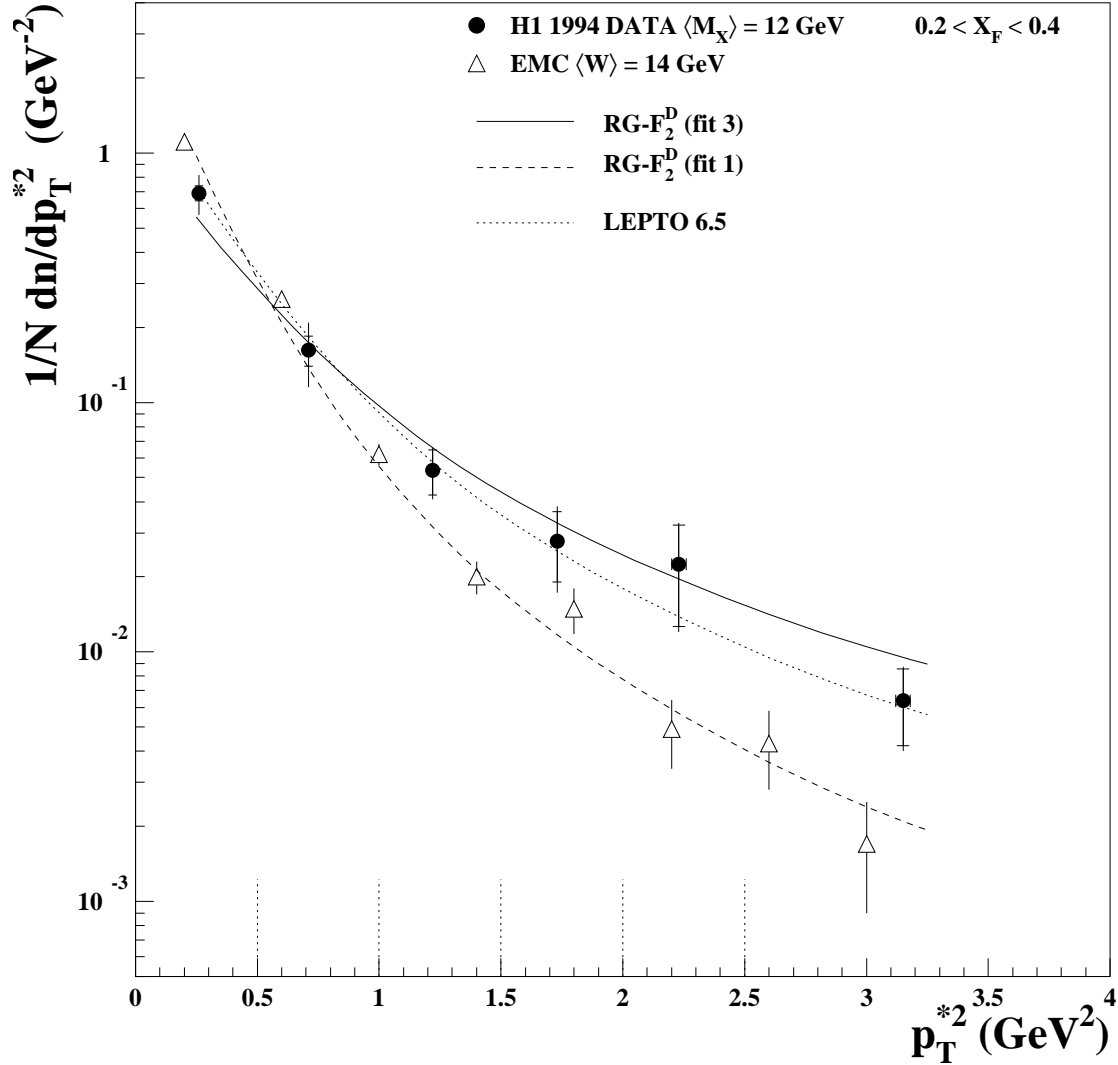,
          width=14cm,bburx=515,bbury=515,bbllx=25,bblly=40}}
    \end{picture}
    \caption{\footnotesize The $p_T^{*2}$ 
             distribution, showing 
             H1 $\gamma^*\pom$ data in the $\gamma^* \pom$ CM frame
             in the kinematic region
             $7.5 < Q^{2} < 100\,$GeV$^2$, 
             $0.05 < y < 0.6$, $\xpom < 0.025$, $8<M_X<18$\,GeV,
             $|t|<1\,\gevsq$, $M_Y<1.6\,{\rm GeV}$,
             and $0.2<x_F<0.4$, together with EMC
             $\mu p$ DIS data in the $\gamma^* p$ CM frame.
             Each point is plotted at the horizontal position where the
             predicted probability density function has its mean value 
             in the bin, as prescribed in~\protect\cite{stick}.
             The uncertainty in the horizontal position
             due to the model-dependence of the shape of the distribution 
             is indicated by horizontal bars, but in most cases is
             contained within the width of the symbol.
             The bin boundaries are marked by vertical dotted lines.
             Also shown are the predictions of the following Monte Carlo 
             models for the H1 data: 
             RAPGAP with the hard-gluon pomeron structure
             taken from fit~3 in~\protect\cite{H1QCD1} (\rgfd);
             RAPGAP with the pomeron structure containing
             only quarks at the starting scale, taken from fit 1 
             in~\protect\cite{H1QCD1} (\rgqq);
             and the soft colour interaction model as implemented in 
             LEPTO~6.5.}
    \label{fig:ptsq}
\end{center}
\end{figure}

The $p_T^{*2}$ distribution, normalised by the number of events $N$, 
is shown in figure~\ref{fig:ptsq}.
Results are shown for the range $0.2<x_F<0.4$.  This restricts the 
data to the current region, where a comparison between H1 and EMC data 
is more meaningful, and matches the $x_F$ range chosen by EMC\@.
The number of high-$p_T^*$ charged particles 
($p_T^{*2}\,\gsim\,2.0$ GeV$^2$) 
is significantly higher in the $\gamma^* \pom$ data than in the
EMC $\gamma^* p$ data.  
This points to a larger contribution from scattering from gluons
in the diffractive case than in inclusive DIS\@.

The $p_T^{*2}$ distribution is well described by the \rgfd\ calculation, 
which predicts a harder spectrum than the \rgqq\ model because of the 
role of the BGF process. The data are also well described by LEPTO~6.5.
The \rgqq\ model does not describe the distribution well at high $p_T^{*2}$.
It should also be noted that the \rgqq\ predictions for the $p_T^{*2}$ 
distribution are in good agreement with the EMC $\gamma^* p$ data.
This supports the argument that the EMC data (at a mean $Q^2$ of
$12\,\gevsq$) are dominated by quarks
and that the higher $p_T^{*2}$ in the diffractive data is not an
effect of the larger $Q^2$ at H1
(where the mean $Q^2$ in this analysis is $25\,\gevsq$).

\begin{figure}[tb] \unitlength 1cm
  \begin{center}
    \begin{picture}(15.,15.)
     \put(0.5,0.5){\epsfig{
         file=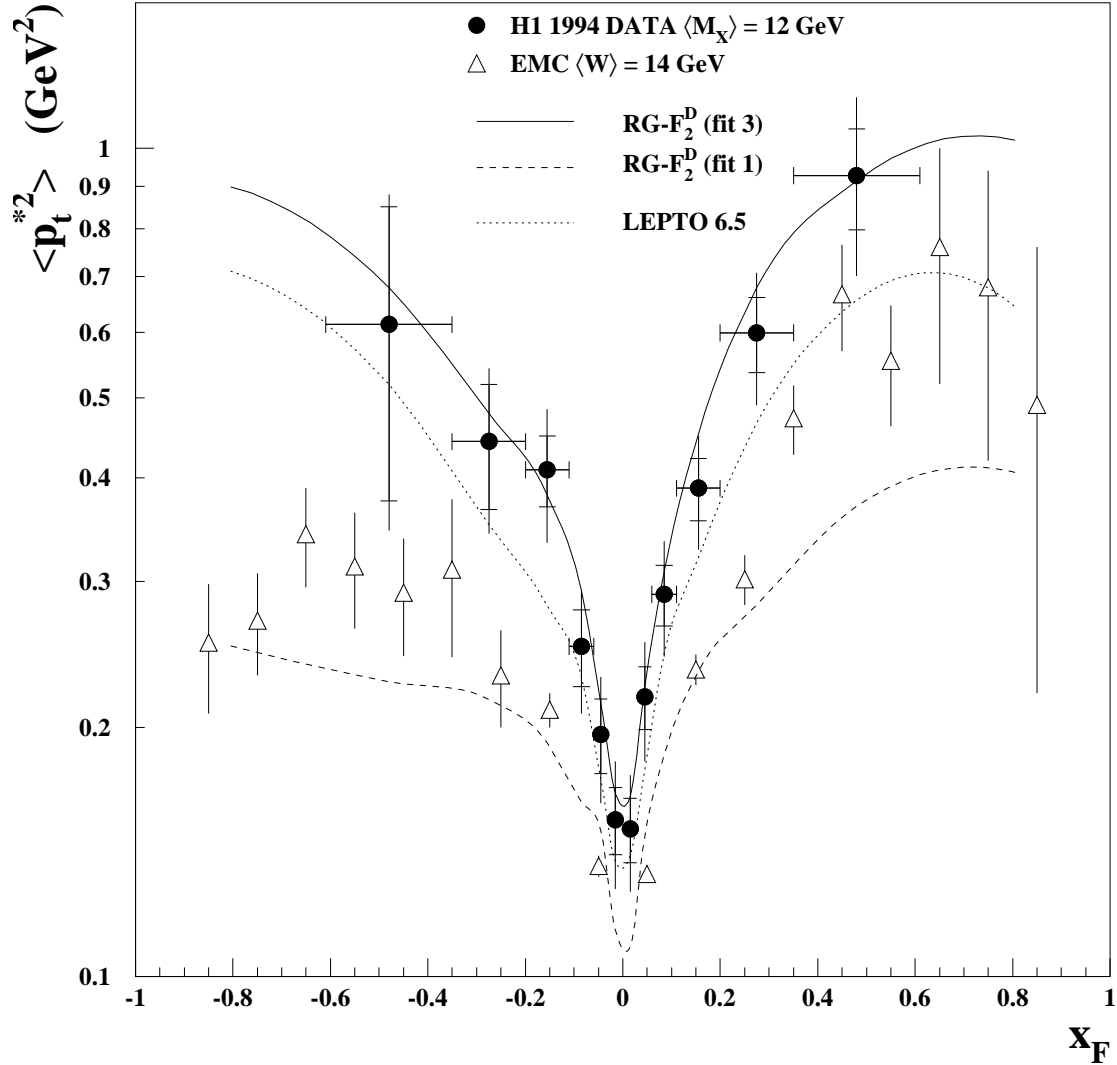,
         width=14cm,bburx=515,bbury=515,bbllx=25,bblly=40}}
    \end{picture}
    \caption{\footnotesize Seagull plot showing 
             H1 $\gamma^*\pom$ data in the $\gamma^* \pom$ CM frame
             in the kinematic region
             $7.5 < Q^{2} < 100\,$GeV$^2$, 
             $0.05 < y < 0.6$, $\xpom < 0.025$,
             $|t|<1\,\gevsq$, $M_Y<1.6\,{\rm GeV}$
             and $8<M_X<18$\,GeV,
             together with EMC
             $\mu p$ DIS data in the $\gamma^* p$ CM frame.
             Positive $x_F$ corresponds to the
             direction of motion of the incoming photon.
             The points are plotted at the centre of the bin in the
             horizontal coordinate, and the horizontal bars indicate the
             width of the bin.
             Also shown are the predictions of the following Monte Carlo 
             models for the H1 data: 
             RAPGAP with the hard-gluon pomeron structure
             taken from fit~3 in~\protect\cite{H1QCD1} (\rgfd);
             RAPGAP with the pomeron structure containing
             only quarks at the starting scale, taken from fit 1 
             in~\protect\cite{H1QCD1} (\rgqq);
             and the soft colour interaction model as implemented in 
             LEPTO~6.5. \vspace*{5mm} }
    \label{fig:seagull}
\end{center}
\end{figure}

Confirmation of the general trends discussed above is obtained 
from the ``seagull plot'', shown in figure~\ref{fig:seagull},
in which the mean transverse momentum squared, $\langle p_T^{*2} \rangle$, 
is plotted as a function of $x_F$\@.
It is observed that $\langle p_T^{*2} \rangle$ is significantly higher 
in the H1 diffractive data than in the EMC $\gamma^* p$ data at all but
the highest $x_F$ values, confirming the conclusions drawn from the 
comparisons in figure~\ref{fig:ptsq}.
The seagull plot also shows a greater degree of symmetry than is
present in the $\gamma^* p$ data (cf.\ figure~\ref{fig:xf}).
As well as the effect of baryon number conservation, this is indicative 
of more parton radiation in the target fragmentation region.
This is consistent with having a more point-like partonic system in 
the target fragmentation hemisphere in the diffractive data than in
the $\gamma^* p$ data, where the extended nature of the proton remnant
leads to a restricted phase space for parton radiation.
Note, however, that there might be an indication of a small asymmetry between
the two hemispheres in the energy-flow distribution 
(figure~\ref{fig:eflow}).

The significant increase in $\langle p_T^{*2} \rangle$ as $|x_F|$ increases 
from $0$ to $\sim 0.5$ is well described by the \rgfd\ calculation
and reasonably well described by LEPTO~6.5,
whereas the $\langle p_T^{*2} \rangle$ values predicted by the 
\rgqq\ model are too low across most of the $x_F$ range, 
in accordance with the above discussion of figure~\ref{fig:ptsq}.

%
%
%
\section{Summary and Conclusions \label{sec_concl}}

Energy-flow and charged-particle spectra have been measured in
diffractive deep-inelastic scattering at HERA in the centre-of-mass (CM)
frame of the photon dissociation ($\gamma^* \pom$) system.
The data support the conclusion reached in the analysis of the
diffractive structure function $F_2^{D(3)}$ that, at low $Q^2$,
the momentum of the diffractive exchange is carried largely by hard gluons.
Thus, significant transverse momentum and energy flow
are produced by the hard subprocess
in the boson-gluon fusion mechanism, and also arise from the
enhanced soft gluon radiation associated with the gluon from the
diffractive exchange.
In the photon dissociation picture, the same data can correspondingly be 
interpreted as evidence for a significant contribution from events 
with photon fluctuation into Fock states with one or more gluons.

The data have been compared with the $\gamma^* p$ data of the
EMC collaboration, with the $\gamma^* p$ CM energy ($W$) 
of the EMC inclusive DIS data similar to the $\gamma^* \pom$ 
CM energy ($M_X$) of the H1 diffractive data.
It is seen that additional transverse momentum is produced in diffractive
scattering compared to the $\gamma^* p$ case. 
A striking contrast is observed in the seagull plot, 
indicating an approximate symmetry between the 
target and current hemispheres in diffractive DIS, as would be expected
in the hard-gluon picture of the pomeron and in the photon dissociation 
picture, whereas radiation is significantly suppressed in the 
proton-remnant region in non-diffractive DIS\@.
 
The features of hadron production are well reproduced by a model
featuring a factorisable pomeron flux in the proton if the partonic
structure of the pomeron is dominated by hard gluons
at the starting scale of $Q_0^2=3\,\gevsq$ used in a DGLAP analysis of the 
diffractive structure function $\fiidiii$.
In contrast, a model with a pomeron consisting, at low $Q^2$, primarily 
of quarks fails to describe the data in several respects.
With respect to the $\gamma^* \pom$ axis, the quark-based pomeron produces 
too little energy flow and particle production in the central 
region, but too much at large $|\eta|$ and $|x_F|$.
In the $p_T^{*2}$ and ``sea\-gull'' distributions,
this model produces too little transverse momentum.
A model based on soft colour interactions (SCI) also gives an acceptable
description of the data.

\section*{Acknowledgments}

We are grateful to the HERA machine group, whose outstanding
efforts have made and continue to make this experiment 
possible. We thank
the engineers and technicians for their work in constructing and 
maintaining the H1 detector, our funding agencies for 
financial support, the
DESY technical staff for their continual assistance, 
and the DESY directorate for the
hospitality which they extend to the 
non-DESY members of the collaboration.


\begin{thebibliography}{99}

\bibitem{rapgap-evts}
ZEUS Collaboration (M. Derrick et al.), 
Phys.\ Lett.\ B315 (1993) 481; \\
H1 Collaboration (T. Ahmed et al.), 
Nucl.\ Phys.\ B429 (1994) 477.

\bibitem{f2d93}
H1 Collaboration (T. Ahmed et al.), 
Phys.\ Lett.\ B348 (1995) 681; \\
ZEUS Collaboration (M. Derrick et al.), 
Z.\ Phys.\ C68 (1995) 569.

\bibitem{H1QCD1}
H1 Collaboration (C. Adloff et al.),
Z. Phys.\ C76 (1997) 613.

\bibitem{DGLAP}
G. Altarelli and G. Parisi,
Nucl.\ Phys.\ B126 (1977) 298; \\
Yu.\ L. Dokshitser,
Sov.\ Phys.\ JETP 46 (1977) 641; \\
V. N. Gribov and L. N. Lipatov,
Sov.\ J.\ Nucl.\ Phys.\ 15 (1972) 438, 675.

\bibitem{zeus:shape}
ZEUS Collaboration (J. Breitweg et al.), 
DESY preprint 97--202,
to be published in Phys.\ Lett.\ B.

\bibitem{h1:thrust}
H1 Collaboration (C. Adloff et al.),
Eur.\ Phys.\ J. C1 (1998) 495.

\bibitem{h1:jets}
P. Marage,
Proceedings of the $5^{th}$ International Workshop on Deep Inelastic
Scattering and QCD (DIS 97), Chicago, April 1997, World Scientific (1997).

\bibitem{phot-diss}
N. N. Nikolaev and B. G. Zakharov,
Z. Phys.\ C53 (1992) 331; \\
A. Hebecker,
Talk given at
Madrid Workshop on Low-$x$ Physics, Madrid, Spain, 18-21 June 1997,
preprint hep--ph/9710475.

\bibitem{ajm}
J. D. Bjorken and J. Kogut,
Phys. Rev. D8 (1973) 1341.

\bibitem{sci}
A. Edin, G. Ingelman and J. Rathsman,
Phys.\ Lett.\ B366 (1996) 371.

\bibitem{mc:lepto}
G. Ingelman, A. Edin and J. Rathsman,
Comp.\ Phys.\ Comm.\ 101 (1997) 108.

\bibitem{H1-det} 
H1 Collaboration (I. Abt et al.), 
Nucl. Instr. and Meth. A368 (1997) 310; \\ 
H1 Collaboration (I. Abt et al.),   
Nucl. Instr. and Meth. A368 (1997) 386.

\bibitem{triggeff}
H1 Collaboration (S. Aid et al.),
Nucl.\ Phys.\ B470 (1996) 3.

\bibitem{thesis:mehta}
A. Mehta,
PhD thesis,
University of Manchester 1994 (unpublished).

\bibitem{h1:diff-photo}
H1 Collaboration (C. Adloff et al.),
Z.\ Phys.\ C74 (1997) 221.

\bibitem{thesis:hfs}
C. M. Cormack,
PhD thesis,
University of Liverpool 1997 (unpublished).

\bibitem{rapgap}
H.\ Jung, 
Comp.\ Phys.\ Comm.\ 86 (1995) 147;\\
H.\ Jung, 
{\em RAPGAP\,2.02 Program Manual}, 1996 (unpublished).

\bibitem{MEPS}
M. Bengtsson and T. Sj\"ostrand,
Z. Phys.\ C37 (1988) 465;\\
M. Bengtsson, G. Ingelman and T. Sj\"ostrand,
Proceedings of the HERA Workshop 1987,
ed.\ R. D. Peccei, DESY, Hamburg, 1988.

\bibitem{JETSET}
T.\ Sj\"ostrand, 
Comp.\ Phys.\ Comm.\ 82 (1994) 74.

\bibitem{ARIADNE}
L.\ L\"{o}nnblad,
Comp.\ Phys.\ Comm.\ 71 (1992) 15.

\bibitem{HERACLES}
A. Kwiatkowski, H. Spiesberger and H.-J. M\"ohring,
Comp.\ Phys.\ Comm.\ 69 (1992) 155.

\bibitem{owens}
J. F. Owens,
Phys.\ Rev.\ D30 (1984) 943.

\bibitem{mrsh}
A. D. Martin, W. J. Stirling and R. G. Roberts,
Phys.\ Rev.\ D50 (1994) 6734.

\bibitem{EMC}
EMC Collaboration (M. Arneodo et al.),
Phys.\ Lett.\ B149 (1984) 415; \\
EMC Collaboration (M. Arneodo et al.), 
Z.\ Phys.\ C35 (1987) 417.

\bibitem{dok:qcd}
Yu.\ L. Dokshitzer,
1995 European School of High-Energy Physics,
ed.\ N. Ellis, M. Neubert, CERN 96--04 (1996) p59.

\bibitem{opal:gluons}
OPAL Collaboration (K. Ackerstaff et al.),
Eur.\ Phys.\ J. C1 (1998) 479.

\bibitem{stick}
G. D. Lafferty and T. R. Wyatt,
Nucl. Instr. and Meth. A355 (1995) 541.


\end{thebibliography}
\end{document}